\documentclass[11pt]{article}
\usepackage{lineno,natbib,amsmath}
\usepackage{amsthm}
\usepackage{here}
\usepackage{lscape}
\usepackage{longtable}
\usepackage{amsfonts,bm,graphicx,graphics}
\usepackage{lipsum} % just for dummy text- not needed for a longtable

\newtheorem*{remark}{Remark}
\newtheorem{proposition}{Proposition}

\usepackage[paperwidth=26cm]{geometry}

\usepackage{multirow}

\usepackage{listings}
\usepackage{color}

\definecolor{codegreen}{rgb}{0,0.6,0}
\definecolor{codegray}{rgb}{0.5,0.5,0.5}
\definecolor{codepurple}{rgb}{0.58,0,0.82}
\definecolor{backcolour}{rgb}{0.95,0.95,0.92}

\lstdefinestyle{mystyle}{
    backgroundcolor=\color{backcolour},
    commentstyle=\color{codegreen},
    keywordstyle=\color{magenta},
    numberstyle=\tiny\color{codegray},
    stringstyle=\color{codepurple},
    basicstyle=\footnotesize,
    breakatwhitespace=false,
    breaklines=true,
    captionpos=b,
    keepspaces=true,
    numbers=left,
    numbersep=5pt,
    showspaces=false,
    showstringspaces=false,
    showtabs=true,
    tabsize=2
}

\title{Extended Poisson INAR(1) processes with equidispersion, underdispersion and overdispersion}

\author{Marcelo Bourguignon$^{1}$\thanks{Corresponding
author: Marcelo Bourguignon. Email: m.p.bourguignon@gmail.com},
Josemar Rodrigues$^{2,**}$, and Manoel Santos-Neto$^{3,4,***}$\\
 $^{1}$Department of Statistics, Federal University of Rio Grande do Norte, Natal, BRA\\
    $^{2}$Department of Mathematics and Statistics, University of S\~ao Paulo, S\~ao Carlos, BRA\\
    $^{3}$Department of Statistics, Federal University of S\~ao Carlos, S\~ao Carlos, BRA\\
    $^{4}$Department of Statistics, Federal University of Campina Grande, Campina Grande, BRA
}

\begin{document}
%\linenumbers
%\listofchanges
\maketitle

\begin{abstract}
Real count data time series often show  the phenomenon of the underdispersion and overdispersion.
In this paper, we develop two extensions of the
first-order integer-valued autoregressive process with Poisson innovations, based on binomial thinning, for modeling integer-valued
time series with equidispersion, underdispersion and overdispersion.
The main properties of the models are derived.
The methods of conditional maximum likelihood, Yule-Walker and conditional least squares are used for
estimating the parameters, and their asymptotic properties are established.
We also use a test based on our processes for checking if the count time series considered is overdispersed or underdispersed.
The proposed models are fitted to time series of number of weekly sales and of cases of family violence illustrating its capabilities
in challenging cases of overdispersed and underdispersed count data.

%\begin{classcode}
%Primary XXX \sep Secondary  XXX.
%\end{classcode}
\end{abstract}

\section{Introduction}
\hskip 0.65cm

\cite{mckenzie85} and \cite{alosh87}, independently,  introduced the integer-valued autoregressive
(\textbf{INAR}) process  $\{X_{t}\}_{t\in\mathbb{Z}}$  with one lag using binomial thinning operator as follows\begin{equation}
\label{eq-1}
X_t=\alpha\circ X_{t-1} + \epsilon_t, \quad t \in \mathbb{Z},
\end{equation}
where $0 \leq \alpha < 1$, $\{\epsilon_t \}_{t\in\mathbb{Z}}$ is a sequence of
independent and identically distributed integer-valued random variables, called
innovations, with $\epsilon_t$ independent of $X_{t-k}$ for all $k \geq 1$,
$\textrm{E}(\epsilon_t)=\mu_{\epsilon}$ and
$\textrm{Var}(\epsilon_t)=\sigma_{\epsilon}^{2}$.  The binomial thinning
operator  $``\circ"$ \citep[][]{steu1979} is defined by
$\alpha \circ X_{t-1} := \sum_{j=1}^{X_{t-1}}W_{j}$, where the  counting series $\{W_{j}\}_{j\geq 1}$ is a sequence of independent and identically distributed Bernoulli
random variables with $\textrm{Pr}(W_j = 1) = 1 - \textrm{Pr}(W_j = 0) = \alpha$. From the results of  \cite{alosh87}, we have that $\alpha \in [0, 1)$ and $\alpha = 1$ are the conditions of (strictly) stationarity and non-stationarity of the process $\{X_{t}\}_{t\in\mathbb{Z}}$, respectively. Also, $\alpha = 0\ (\alpha > 0)$ implies the independence (dependence) of the observations of $\{X_{t}\}_{t\in\mathbb{Z}}$.

The practical motivation for this type of process is the need to model count series with correlated observations. Some examples are daily counts of epileptic seizure in one patient, the number of generics in the pharmaceutical market, the number of guest nights in hotels, the number of different IP addresses, the monthly number of active customers of a mobile phone service provider, daily number of traded stocks in a firm, daily number of visitors to a website, monthly incidence of a disease, and so on. For more details see, for example, \cite{Hellstrom:2001aa, Brannas:2002aa, Weis:2007aa, Weis:2009aa, Barreto-Souza:2015aa, Bourguignon:2016aa} and \cite{Bourguignon:2017aa}.

If $\varphi_{X}(s)$ and $\varphi_{\epsilon}(s)$ denote the probability
generating function (pgf) of $\{X_{t}\}_{t\in\mathbb{Z}}$ and $\{\epsilon_t \}_{t\in\mathbb{Z}}$, respectively, then
the stationary marginal distribution of $\{X_{t}\}_{t\in\mathbb{Z}}$ can be determined from the equation
$\varphi_{X}(s) = \varphi_{X}(1 - \alpha(1-s)) \cdot \varphi_{\epsilon}(s)$,
which allows for various types of marginal distributions, including the Poisson~\citep[][]{alosh87}, geometric~\citep[][]{Al-Osh:1988aa}, generalized Poisson~\citep[][]{Alzaid:1993aa} and Poisson-geometric ~\citep[][]{Bourguignon:2016ab} distributions.
Also, the marginal distribution of model (\ref{eq-1}) may be expressed in terms of arrival process $\{\epsilon_t \}_{t\in\mathbb{Z}}$~\citep[][]{alosh87} as $X_t \stackrel{d}{=} \sum_{i=0}^{\infty}\alpha^i \circ \epsilon_{t-i}$. Thus, the pgf of $\{X_{t}\}_{t\in\mathbb{Z}}$ is given by
$$\varphi_{X}(s) = \prod\limits_{i=0}^{\infty}\varphi_{\epsilon}(1-\alpha^i + \alpha^is).$$

The \textbf{INAR}(1) process is a homogeneous Markov chain and the 1-step transition probabilities of this process are given by
\begin{equation*}\label{ml}
\textrm{Pr}(X_t = k|X_{t-1} = l) =
\sum_{i=0}^{\min(k,l)}\textrm{Pr}(B_l^{\alpha} = i)\cdot\textrm{Pr}(\epsilon_t = k-i), \quad k, l \geq 0,
\end{equation*}
where $B_n^\alpha \sim \textrm{Binomial}(\alpha, n)$ with $\alpha \in (0, 1)$ and $n \in \mathbb{N}$.
The mean and variance of $\{X_{t}\}_{t\in\mathbb{Z}}$ are given by
\begin{equation*}
\mu_X:=\textrm{E}(X_{t}) =  \frac{\mu_{\epsilon}}{1-\alpha} \quad \textrm{and} \quad \sigma^2_X:=\textrm{Var}(X_{t}) =
\frac{\alpha\,\mu_{\epsilon} + \sigma_{\epsilon}^2}{1-\alpha^2},
\end{equation*}
respectively.
A commonly used variability measure of a random variable is the Fisher index of dispersion, defined by $\textbf{FI}_X = \textbf{Var}(X)/\textbf{E}(X)$, that is a measure of aggregation or disaggregation, for more details see~\cite{johnson2005}, pag. 163. Thus, the Fisher index of dispersion of $\{X_{t}\}_{t\in\mathbb{Z}}$  in (\ref{eq-1}) is given by
\begin{equation}\label{index}
\textbf{FI}_X= \frac{\textbf{FI}_\epsilon + \alpha}{1 + \alpha},
\end{equation}
where $\textbf{FI}_\epsilon$ is the Fisher index of dispersion of
the innovations $\{\epsilon_t\}_{t \in \mathbb{Z}}$.
Furthermore, the autocorrelation function (ACF) at lag $h$ is given by $\rho_X(h) = \alpha^h$, $h \geq 0$, i.e., it is of \textbf{AR}(1)-type, but only positive autocorrelation is allowed.

The Equation (\ref{index}) shows that the dispersion behaviour of the observations $\{X_{t}\}_{t\in\mathbb{Z}}$ is controlled
by that of the innovations $\{\epsilon_t\}_{t \in \mathbb{Z}}$, i.e., the Equation (\ref{index}) implies that we obtain an \textbf{INAR}(1)
process with the distribution of the $\{X_{t}\}_{t\in\mathbb{Z}}$ being equidispersion, underdispersion or overdispersion iff the distribution of the innovations is chosen to be equidispersion, underdispersion or overdispersion, respectively.
%andwe obtain an \textbf{INAR}(1) process with the distribution of the
%$\{X_{t}\}_{t\in\mathbb{Z}}$ being underdispersed iff the distribution of the $\{\epsilon_t\}_{t \in \mathbb{Z}}$ is chosen to be underdispersed.
Thus, a simple approach is to only change the innovations distribution in such a way that the marginal distribution of the process is
underdispersed or overdispersed.

In this context, based on binomial thinning operator,
\cite{Jazi:2012aa} introduced the \textbf{INAR}(1) process with geometric innovations.
\cite{Jazi:2012ab} discussed an \textbf{INAR}(1) process with zero-inflated Poisson innovations.
\cite{Weis:2013aa} studied a stationary INAR(1) process with Good and certain types of power law
weighted Poisson innovations. However, the applications focus on underdispersion.
\cite{Schweer:2014aa} introduced a first-order non-negative integer-valued
autoregressive process with compound Poisson innovations.
\cite{Bourguignon:2015aa} studied a new stationary \textbf{INAR}(1) process with power series innovations.
\cite{Andersson:2014aa} used the signed binomial thinning operator to define a first-order
process with Skellam-distributed innovations. \cite{Fernandez-Fontelo:2017aa} introduced a generalization of the classical Poisson-based \textbf{INAR} models whose innovations follow a Hermite distribution. \cite{Kim:2017aa} considers the \textbf{INAR}(1) process with Katz family innovations. %\cite{Bourguignon:2017aa}
%\replaced[id=Manoel]{Penso que a referencia estah fora de contexto, pois todas as outras fazem referencia
%a diferentes ditribuiccoes usadas para a inovaccao}{proposed an INAR(1) process with
%BerG marginals, which can be used for modeling time series of counts with equidispersion,
%underdispersion and overdispersion. However, the parameters restrictions aren't liberal.}
This paper aims to give a contribution in this direction.

The main objective of this paper is to propose two new  binomial thinning \textbf{INAR}(1) processes
with double Poisson and generalized Poisson innovations, denoted by \textbf{INARDP}(1) and \textbf{INARGP}(1), respectively, for modeling nonnegative integer-valued time series with equidispersion, underdispersion or overdispersion.
We are choosing these distributions (double Poisson and generalized Poisson) for the $\{\epsilon_t\}_{t \in \mathbb{Z}}$
as ideal option among other candidates with overdispersion or underdispersion relative to Poisson distribution,
because, the other options %(Poisson, geometric, Poisson mixture, Good, compound Poisson, Poisson-Geometric, mixed Poisson)
have only one type of dispersion or the probability mass function of distribution and moments of innovations are very complicated. We also propose a test [using the test provided by \cite{Schweer:2016aa}] based on our models for checking
if the count time series considered is overdispersed or underdispersed. Additionally, we will provide a comprehensive account of the mathematical
properties of these two new processes which are very easy to obtain \citep[][]{Eduarda-Da-Silva:2004aa} and the parameter restrictions are liberal $[\alpha \in (0, 1)]$.
Furthermore, the proposed processes have, as a particular case, the Poisson \textbf{INAR}(1) [INARP(1)] process \citep{alosh87}.

%In a very recent paper, \cite{Bourguignon:2017aa} proposed an \textbf{INAR}(1)
%process with BerG marginals, which can be used for modeling time series of counts with equidispersion, underdispersion and overdispersion.
%However, the parameter restrictions aren't liberal.

The article is organized as follows. In Section 2, we construct two new models to properly
capture different types of dispersions, and some of its
properties are outlined. Section 3 discusses some simulation results for the estimation methods.
Two applications with the real data sets are presented in Section 4.

\section{Extended Poisson \textbf{INAR}(1) processes}
\hskip 0.65cm

In this section, we propose two extensions of the Poisson \textbf{INAR}(1) process in
(\ref{eq-1}) to deal with equidispersion, underdispersion and overdispersion problems.

\subsection{Double Poisson \textit{INAR}(1) model}
\hskip 0.65cm

\cite{efron86} proposed, based on the double exponential family, the double Poisson (DP) distribution. This model is indexed
by two parameters $\mu > 0$ and $\phi > 0$. The probability mass function (pmf) is given by
\begin{equation}\label{eq:dp}
\textrm{Pr}(Y = y) = \textrm{Z}(\mu,\phi) \, \left(\sqrt{\phi} \right) \textrm{e}^{-\phi \,\mu}\left[\frac{\textrm{e}^{-y} y^y}{y!} \right]\left( \frac{ {\rm e}\, \mu}{y} \right)^{\phi\, y}, \quad y = 0, 1, 2, \ldots,
\end{equation}
where $\textrm{Z}(\mu,\phi)^{-1} \approx 1 + \frac{1-\phi}{12\mu\phi}\left(1 + \frac{1}{\mu\phi} \right) $. The expected value and the variance are given by
\begin{equation*}\label{eq:mvdp}
\textrm{E}(Y) \approx \mu \quad \mbox{and} \quad \textrm{Var}(Y) \approx \mu/\phi.
\end{equation*}

Thus, the DP distribution allows for both overdispersion ($\phi < 1$) and underdispersion ($\phi > 1$). If $\phi = 1$, the DP distribution collapses to the Poisson distribution. It is possible to show that the pgf is given by
$$\varphi_Y(s) =  \frac{\textrm{Z}(\mu \,s, \phi)}{\textrm{Z}(\mu, \phi)}.$$

Now, let $\{\epsilon_t\}_{t \in \mathbb{Z}}$ be a sequence of discrete i.i.d. random variables following a DP
distribution  with probability mass function given in (\ref{eq:dp}).
In short, we name this process as the \textrm{INARDP}(1) process.
Thus,
the transition probabilities of this process are given by
\begin{equation}\label{cmldp}
\textrm{Pr}(X_t = k|X_{t-1} = l) =
 Z(\mu,\phi)\,\sqrt{\phi}\,{\rm e}^{-\phi \mu}\,\sum\limits_{i=0}^{\textrm{min}(k,l)}{l \choose i}\alpha^{i}(1-\alpha)^{l-i}\,
\frac{{\rm e}^{-(k-i)} (k-i)^{(k-i)}}{(k-i)!} \cdot\left(\frac{ {\rm e}\, \mu}{k-i}\right)^{\phi (k-i)}.
\end{equation}

The mean, variance and Fisher index of dispersion of $\{X_t\}_{t\in \mathbb{Z}}$ are given by
\begin{equation*}\label{momentsgeo}
 \mu_X = \frac{\mu}{(1-\alpha)}, \quad
  \sigma^{2}_X = \frac{\mu(1+\alpha\,\phi)}{\phi(1-\alpha^2)}
\quad \textrm{and} \quad
\textrm{FI}_X = \frac{1 + \alpha\,\phi}{\phi +  \alpha\,\phi}.
\end{equation*}

It follows that this process shows equidispersion for $\phi = 1$, while we have underdispersion for $\phi > 1$, and
overdispersion for $\phi < 1$. Note that the parameter $\mu$ does not change the dispersion index
of the process. Table \ref{tabledp} contains the dispersion index of
the \textrm{INARGP}(1) model for various parameter values.

\begin{table}[!htb]
\begin{center}\caption{Dispersion index of \textrm{INARDP}(1) model for various values of $\alpha$ and $\phi$.}\label{tabledp}
\begin{tabular}{llccc}
\hline
$\phi \downarrow$ & $\alpha \rightarrow$  &  $\alpha = 0.3$&$\alpha = 0.5$ &$\alpha = 0.7$\\ \hline
$\phi = 0.3$    &                       &  2.7949              &  2.5556           & 2.3725             \\
$\phi = 0.5$    &                       &  1.7692              &  1.6667           & 1.5882          \\
$\phi = 0.7$    &                       &  1.3297               &  1.2857           & 1.2521             \\
$\phi = 1.3$    &                       &  0.8225              &  0.8462           &  0.8642            \\
$\phi = 1.5$    &                       &  0.7436              & 0.7778            &  0.8039         \\
$\phi = 1.7$    &                       &  0.6833              & 0.7255            &  0.7578            \\
\hline
\end{tabular}
\end{center}
\end{table}

The conditional expectation and the conditional variance are given, respectively, by
$$\textrm{E}(X_t|X_{t-1}) = \alpha \, X_{t-1} + \mu \quad
\textrm{and} \quad
\textrm{Var}(X_t|X_{t-1}) = \alpha(1-\alpha)X_{t-1} + \mu/\phi.$$

In practice, the true values of the model parameters of the process are not known but have to be estimated from a given realization $X_1, \ldots, X_T$ of the process.
We consider three estimation methods, namely, conditional least squares, Yule–Walker
and conditional maximum likelihood.

\subsubsection{Conditional least squares estimation}
\hskip 0.65cm

The conditional least squares (CLS) estimator $\bm{\widehat\eta} = (\widehat{\alpha}_{\mathrm{CLS}}, \widehat{\mu}_{\mathrm{CLS}}, \widehat{\phi}_{\mathrm{CLS}})^{\mathsf{T}}$
of $\bm{\eta} = (\alpha, \mu, \phi)^{\mathsf{T}}$ is given by
$$\widehat{\bm{\eta}} = \textrm{arg}\min_{\bm{\eta}}(S_{T}(\bm{\eta})),$$
where $S_{T}(\bm{\eta}) = \sum_{t=2}^{T}[X_t - \alpha\,X_{t-1} - \mu]^2$.
Then, the CLS estimators of $\alpha$ ($\widehat{\alpha}_{\mathrm{CLS}}$)  , and $\mu$ ($\widehat{\mu}_{\textrm{CLS}}$), can be written in closed form as
\begin{eqnarray}\label{clsPD1}
\widehat{\alpha}_{\mathrm{CLS}} =
\frac{(T-1)\sum_{t=2}^{T}X_{t}X_{t-1}-\sum_{t=2}^{T}X_t\sum_{t=2}^{T}X_{t-1}}{(T-1)\sum_{t=2}^{T}X_{t-1}^{2}
- \left(\sum_{t=2}^{T}X_{t-1}\right)^2}
\quad
\textrm{and}
\quad
\widehat{\mu}_{\textrm{CLS}} =  \frac{\sum_{t=2}^{T}X_t-\widehat{\alpha}_{\mathrm{CLS}}\sum_{t=2}^{T}X_{t-1}}{T-1}.
\end{eqnarray}

\begin{proposition}\label{asymptotic1}
The estimators $\widehat{\alpha}_{\textrm{CLS}}$ and $\widehat{\mu}_{\textrm{CLS}}$ given in (\ref{clsPD1}) are strongly consistent and satisfy the asymptotic normality
$$\sqrt{T}[(\widehat{\alpha}_{\textrm{CLS}},\widehat{\mu}_{\textrm{CLS}})^\top-(\alpha,\mu)^\top]\stackrel{d}{\longrightarrow} \textrm{N}_2((0,0)^\top,\bm{\Sigma}),$$
as $T\rightarrow\infty$, where the asymptotic covariance matrix $\bm{\Sigma}$ is given by
\begin{eqnarray*}\label{G1}
\bm{\Sigma} =
\left(
\begin{array}{ccccc}
\frac{\gamma\,\alpha(1-\alpha)^3(1+\alpha)^2\phi^2}{\mu^2(1+\alpha\,\phi)^2} + 1- \alpha^2 &&&&\alpha(1-\alpha) - \mu(1+\alpha) - \frac{\gamma\,\alpha(1-\alpha^2)^2\phi^2}{\mu(1+\alpha\,\phi)^2}  \\ \\ \\
\alpha(1-\alpha) - \mu(1+\alpha) - \frac{\gamma\,\alpha(1-\alpha^2)^2\phi^2}{\mu(1+\alpha\,\phi)^2}&&&&
\frac{\gamma\,\alpha(1-\alpha)(1+\alpha)^2\phi^2}{(1+\alpha\,\phi)^2}+\mu\cdot\frac{1+\alpha}{1-\alpha}+\frac{\mu}{\phi}-\alpha\,\mu
\end{array}
\right),
\end{eqnarray*}
with $\gamma = \mu_{X,3} - 3\,\mu_X\cdot\sigma^2 - \mu_X^3, \mu_{X,3} = \textrm{E}(X_t^3)$.
\end{proposition}

For estimation of the parameter $\phi$, we will use the two-step CLS estimation methods proposed by \cite{Karlsen:1988aa} which consists in  minimizing the function
\begin{eqnarray*}
Q_T(\bm{\eta}) = \sum_{t=2}^{T}\left\{[X_t - \textrm{E}(X_t|X_{t-1})]^2 -\textrm{Var}(X_t|X_{t-1})\right\}^2
= \sum_{t=2}^{T}\left[(X_t - \alpha \,X_{t-1} - \mu)^2 -\alpha(1-\alpha)X_{t-1} + \mu/\phi)\right]^2.
\end{eqnarray*}
Thus, solving the equation $\partial Q_T(\bm{\eta})/\partial \phi = 0$ and replacing $\alpha$ and $\mu$ with the appropriate
CLS estimates, we obtain that the CLS estimate of
the parameter $\phi$ is given by
\begin{eqnarray*}\label{clsPD3}
\widehat{\phi}_{\mathrm{CLS}} =  \frac{\sum_{t=2}^{T}X_t-\widehat{\alpha}_{\mathrm{CLS}}\sum_{t=2}^{T}X_{t-1}}{\sum_{t=2}^{T}\left[(X_t - \widehat{\alpha}_{\mathrm{CLS}}X_{t-1}
- \widehat{\mu}_{\mathrm{CLS}})^2 - \widehat{\alpha}_{\mathrm{CLS}}(1-\widehat{\alpha}_{\mathrm{CLS}})X_{t-1}\right]}.
\end{eqnarray*}

\subsubsection{Yule-Walker estimation}
\hskip 0.65cm

Addtionally,  we propose the method of moments based on the sample quantities of $\rho_X(1), \textbf{E}(X_t)$ and $\textbf{Var}(X_t)$ whose  estimators of $\alpha, \mu$ and $\phi$ are given by
\begin{equation*}\label{YWDP}
\widehat{\alpha}_{\mathrm{YW}} =
\frac{\sum\limits_{t=1}^{T-1}(X_t - \overline{X})(X_{t+1} -
\overline{X})}{\sum_{t=1}^{T}(X_t - \overline{X})^2}, \quad
\widehat{\mu}_{\mathrm{YW}} = (1-\widehat{\alpha}_{\mathrm{YW}})\overline{X}
\quad
\textrm{and}
\quad
\widehat{\phi}_{\mathrm{YW}} = \frac{\overline{X}}{\widehat\gamma(0)(1+\widehat{\alpha}_{\mathrm{YW}})
- \overline{X}\,\widehat{\alpha}_{\mathrm{YW}}},
\end{equation*}
respectively, where $\overline{X} = (1/T)\sum\limits_{t=1}^{T}X_t$.

\subsection{Generalized Poisson \textit{INAR}(1) model}
\hskip 0.65cm

A random variable $Y$ is said to have a generalized Poisson (GP) distribution~\citep[][]{Consul73} if its pmf is given by
\begin{equation}\label{eq:gpd}
\textrm{Pr}(Y = y) =\frac{ \mu (\mu + y \, \phi)^{y-1} {\rm e}^{-(\mu + y \, \phi)}}{y!}, \quad y = 0, 1, \ldots,
\end{equation}
where $\mu > 0$, $|\phi| < 1$ and $\textrm{Pr}(Y = y) = 0$, for $y \geq m$ if $\mu + m \, \phi \leq 0$. It is obvious that that Poisson distribution with parameter $\mu$ is a special case when $\phi = 0$. \cite{Joe:2005aa} proved
that the GP distribution is a mixture of Poisson distribution.  Then, for $0<\phi <1$, the corresponding pgf~\citep[see][]{consul89} is given by
\[
\varphi_Y(s) = \textrm{e}^{\mu[u(s,\phi)-1]},
\]
 where $u(s, \phi)$ is the smaller root of the equation $u = s \, \textrm{e}^{\phi(u-1)}$. The mean and variance are given by
\begin{equation*}\label{eq:momgp}
\textrm{E}(Y) = \frac{\mu}{(1-\phi)}\quad \mbox{and} \quad \textrm{Var}(Y) = \frac{\mu}{(1-\phi)^3},
\end{equation*}
and the third non central moment is
\[
\textrm{E}(Y^3) = \frac{\mu}{(1-\phi)^3}\left[\mu^2(1-\phi)^2  + 3\mu(1-\phi) - 2(1-\phi)+3\right].
\]

The $\textrm{FI}_Y$ is $1/(1-\phi)^2$, soon when $\phi \in (0,1)$ the GP distribution possesses the property of overdispersion. For $\phi \in (-1,0)$ the model possesses the property underdispersion.

Let $\{\epsilon_t\}_{t \in \mathbb{Z}}$ be a sequence of discrete i.i.d. random variables following a GP
distribution with parameters $\mu$ and $\phi$ with probability mass function given in (\ref{eq:gpd}).
In short, we name this process as the \textrm{INARGP}(1) process.
\cite{Alzaid:1993aa} studied an \textrm{INAR(1)} process with GP marginals, but the authors just consider the overdispersion case.

The mean, variance and the dispersion index of $\{X_{t}\}_{t\in\mathbb{Z}}$ are given by
\begin{equation*}
\mu_X =  \frac{\mu}{(1-\alpha)(1-\phi)}, \quad
 \sigma^2_X = \frac{\mu[1 +  \alpha(1-\phi)^2]}{(1-\alpha^2)(1-\phi)^3},
\quad \textrm{and} \quad
\textrm{FI}_X= \frac{1 +  \alpha(1-\phi)^2}{(1+\alpha)(1-\phi)^2}.
\end{equation*}
So, it follows that this model presents equidispersion when $\phi = 0$; underdispersion when
$\phi < 0$, and overdispersion when $\phi > 0$. Note that the parameter $\mu$ does not change the dispersion index
of the process. Table \ref{tablegp} contains the dispersion index of
the \textrm{INARGP}(1) model for various parameter values. Table \ref{tablegp} shows that the \textrm{INARGP}(1) process can capture a small underdispersion.

\begin{table}[!htb]
\begin{center}\caption{Dispersion index of \textrm{INARGP}(1) model for various values of $\alpha$ and $\phi$.}\label{tablegp}
\begin{tabular}{llccc}
\hline
$\phi \downarrow$ & $\alpha \rightarrow$  &  $\alpha = 0.3$&$\alpha = 0.5$ &$\alpha = 0.7$\\ \hline
$\phi = -0.7$    &                        &   0.4969     & 0.5640            &  0.6153            \\
$\phi = -0.5$    &                        &   0.5726    & 0.6296            &  0.6732         \\
$\phi = -0.3$    &                        &   0.6859    & 0.7278            &  0.7598            \\
$\phi = \, \, \, \, 0.3$    &             &   1.8006     & 1.6939             &  1.6122            \\
$\phi = \, \, \, \, 0.5$    &             &   3.3077     & 3                    &  2.7647         \\
$\phi = \, \, \, \, 0.7$    &             &   8.7778     & 7.7407             &  6.9477            \\
\hline
\end{tabular}
\end{center}
\end{table}

The conditional expectation, the conditional variance and the transition probabilities of this process are given by
$$\textrm{E}(X_t|X_{t-1}) = \alpha \, X_{t-1} + \mu/(1-\phi),\quad
 \quad
\textrm{Var}(X_t|X_{t-1}) = \alpha(1-\alpha)X_{t-1} + \mu/(1-\phi)^3\quad \text{and}$$
\begin{equation}\label{cmlgp}
\textrm{Pr}(X_t = k|X_{t-1} = l) =
\sum_{i=0}^{\min(k,l)}{l \choose  i}\alpha^{i}(1-\alpha)^{l-i}\frac{ \mu [\mu + (k-i) \phi]^{k-i-1} {\rm e}^{-[\mu + (k-i) \phi]}}{(k-i)!},
\end{equation}
respectively.

Now, we introduce some properties and we will estimate the unknown parameters $\alpha$, $\mu$ and $\phi$ of the \textrm{INARGP}(1) process. The Yule-Walker estimates are briefly discussed. Also, conditional least squares estimators will be derived, and their asymptotic properties will be considered.

\subsubsection{Yule-Walker estimation}
\hskip 0.65cm

The Yule-Walker (YW) estimators of $\alpha, \mu$ and $\phi$ are based upon the sample autocorrelation
 function $\widehat{\rho}(k)$ with  $\rho_X(1) = \alpha$ and the first moment and the dispersion index of $X_{t}$ given by
$\textrm{E}(X_t) = \mu/[(1 - \alpha)(1 - \phi)]$, and $\textrm{FI}_X=  [1 +  \alpha(1-\phi)^2]/[(1+\alpha)(1-\phi)^2]$, respectively.
Let $X_1, X_2, \ldots, X_T$ be a random sample of size $T$ from the \textbf{INARPG}(1) process. Then, the YW estimators of $\alpha, \mu$ and $\phi$
are given by
\[
\widehat{\alpha}_{\mathrm{YW}} =
\frac{\sum\limits_{t=1}^{T-1}(X_t - \overline{X})(X_{t+1} -
\overline{X})}{\sum_{t=1}^{T}(X_t - \overline{X})^2}, \quad
\widehat{\mu}_{\mathrm{YW}} = (1-\widehat{\alpha}_{\mathrm{YW}})(1-\widehat{\phi}_{\mathrm{YW}})\overline{X},
\]
and
\[
\widehat{\phi}_{\mathrm{YW}} = \frac{\widehat{\alpha}_{\mathrm{YW}}\cdot\widehat{\textrm{FI}}_X
- \sqrt{\widehat{\alpha}_{\mathrm{YW}}\cdot\widehat{\textrm{FI}}_X - \widehat\alpha_{\mathrm{YW}} + \widehat{\textrm{FI}}_X }
 - \widehat{\alpha}_{\mathrm{YW}} + \widehat{\textrm{FI}}_X}{\widehat{\alpha}_{\mathrm{YW}}\cdot\widehat{\textrm{FI}}_X - \widehat{\alpha}_{\mathrm{YW}} + \widehat{\textrm{FI}}_X}.
\]

\subsubsection{Conditional least squares estimation}
\hskip 0.65cm

The conditional least squares (CLS) estimator $\bm{\widehat\eta} = (\widehat{\alpha}_{\mathrm{CLS}}, \widehat{\mu}_{\mathrm{CLS}}, \widehat{\phi}_{\mathrm{CLS}})^{\mathsf{T}}$
of $\bm{\eta} = (\alpha, \mu, \phi)^{\mathsf{T}}$ is given by
$$\widehat{\bm{\eta}} = \textrm{arg}\min_{\bm{\eta}}(S_{T}(\bm{\eta})),$$
where $S_{T}(\bm{\eta}) = \sum_{t=2}^{T}[X_t - g(\bm{\eta},X_{t-1})]^2$ and
$g(\bm{\eta},X_{t-1}) = \textrm{E}(X_t|X_{t-1})=\alpha\,X_{t-1} + \mu/(1-\phi)$.
However, note that $\alpha\,X_{t-1} +  \mu/(1-\phi)$ depends on $\mu$ and $\phi$ only through $\mu/(1-\phi)$ implying that it is not possible
to obtain the estimators of $\mu$ and $\phi$.
Thus, we use the CLS method to find estimators for $\alpha$ and $\phi$ assuming that $\mu$ is known.
Then, in this case the CLS estimators of $\alpha$ and $\phi$ can be written in closed form as
\begin{eqnarray}\label{cls1}
\widehat{\alpha}_{\mathrm{CLS}} =
\frac{(T-1)\sum_{t=2}^{T}X_{t}X_{t-1}-\sum_{t=2}^{T}X_t\sum_{t=2}^{T}X_{t-1}}{(T-1)\sum_{t=2}^{T}X_{t-1}^{2}
- \left(\sum_{t=2}^{T}X_{t-1}\right)^2}
\quad
\textrm{and}
\quad
\widehat{\phi}_{\mathrm{CLS}} = 1 - \frac{\mu(T-1)}{\sum_{t=2}^{T}X_t-\widehat{\alpha}\sum_{t=2}^{T}X_{t-1}},
\end{eqnarray}
where $\mu$ will be replaced by some consistent estimator $\widehat{\mu}$ as the one in the previous subsection  given by $\widehat{\mu}_{\mathrm{CLS}}
 = \widehat{\mu}_{\mathrm{YW}}.$

\begin{proposition}\label{asymptotic2}
The estimators $\widehat{\alpha}_{\mathrm{CLS}}$ and $\widehat{\phi}_{\mathrm{CLS}}$ given in (\ref{cls1}) are strongly consistent for
estimating $\alpha$ and $\phi$, respectively, and satisfy the asymptotic normality
$$\sqrt{T}[(\widehat{\alpha}_{\mathrm{CLS}},\widehat{\phi}_{\mathrm{CLS}})^\top-(\alpha,\phi)^\top]\stackrel{d}{\longrightarrow} \textrm{N}_2((0,0)^\top,\bm{V^{-1}WV^{-1}}),$$
where
\begin{eqnarray*}\label{G2}
{\bf V^{-1}} =
\left(
\begin{array}{ccccc}
\frac{\mu_{\epsilon}\sigma_{\epsilon}^{2}}{\mu_{\epsilon}\sigma_{\epsilon}^{2}\mu_{X,2} - (1-\alpha)^2\mu_{X}^4} &&&&-\frac{(1-\alpha)\mu_{X}^2}{\mu_{\epsilon}\sigma_{\epsilon}^{2}\mu_{X,2} - (1-\alpha)^2\mu_{X}^4}  \\ \\ \\
-\frac{(1-\alpha)\mu_{X}^2}{\mu_{\epsilon}\sigma_{\epsilon}^{2}\mu_{X,2} - (1-\alpha)^2\mu_{X}^4}&&&&
\frac{\mu_{X,2}}{\mu_{\epsilon}\sigma_{\epsilon}^{2}\mu_{X,2} - (1-\alpha)^2\mu_{X}^4}
\end{array}
\right)
\end{eqnarray*}
and
\begin{eqnarray*}\label{G3}
{\bf W} =
\left(
\begin{array}{ccccc}
\alpha(1-\alpha)\mu_{X,3} + \sigma_{\epsilon}^{2}\mu_{X,2} &&&& \mu_{\epsilon}\,\sigma_{\epsilon}^{2}(\alpha\,\mu_{\epsilon} + \sigma_{\epsilon}^{2}) \\ \\ \\
\mu_{\epsilon}\,\sigma_{\epsilon}^{2}(\alpha\,\mu_{\epsilon} + \sigma_{\epsilon}^{2})&&&&
\lambda\left[\alpha(1-\alpha)\mu_{X,2} + \mu_X\sigma_{\epsilon}^{2}\right]/(1-\theta)^2
\end{array}
\right),
\end{eqnarray*}
with $\mu_{X,2} = \sigma^2_X + \mu_X^2$ and $\mu_{X,3} = \textrm{E}(X_t^3)$.
\end{proposition}

\begin{remark}({\tt tsinteger} R package) The theoretical results of this paper has been implemented into a piece of statistical software: the {\tt tsinteger} package for {\tt R}~\citep{r:2017, tsinteger}. To install this package, the {\tt R} code below must be used.

\begin{lstlisting}[language=R]
devtools::install_github("projecttsinteger/tsintegerpackage")
\end{lstlisting}

This package contains a collection of utilities for analyzing data from \textbf{INAR}(1) processes. Some of the functions are: {\tt epoinar()}, {\tt est.inar()}, {\tt epoinar.sim()} and {\tt equi.test()}.
\end{remark}

\section{Experimental evaluation}  \label{sec4}
\hskip 0.65cm

This section contains results from a simulation study that illustrates the performances of the different methods of estimation for parameters of the models described in the previous sections. The simulation study was carried out to compare the estimates obtained from the YW, CLS and conditional maximum likelihood (CML) methods. These three methods of estimation was based on their empirical bias and mean square error (MSE).

In our simulation study, a random sample of size $T = 100, 200, 400$ and $800$ was generated and values of $X_{1}$
were independently drawn from the double Poisson (or generalized Poisson) with corresponding values of $X_t$ given by
\[
X_t = 0.3 \circ X_{t-1} + \epsilon_t, %\quad t \in \mathbb{Z},}
\]
where $\epsilon_t$ were set to be independently drawn from the following two studied distributions: a) double Poisson  with parameters $\mu=5.0$ and $\phi = 0.5$ and  $2.0$; and b) generalized Poisson with parameters $\mu=1.0$ and $\phi = -0.5$ and $0.5$. The simulation process was replicated 5,000 times. The empirical means and mean square error of the three methods of estimation were then computed. All simulations were accomplished by using {\tt R} software.
To simulate the innovations process, we simulate a DP and a
GP one with R's {\tt rdpois} ({\tt rmutil} package)
and {\tt rgenpois} ({\tt HMMpa} package) functions, respectively.

\begin{remark}(Conditional maximum likelihood approach)
Let $X_1, X_2, \ldots, X_T$, with $X_1$ fixed, be a random sample of size $T$ from a stationary \textrm{INARDP}(1)  or \textrm{INARGP}(1) process
with vector parameters $\bm{\eta}$. The conditional log-likelihood function for the
\textrm{INARDP}(1)  or \textrm{INARGP}(1) process is given by
\begin{equation*}\label{CML}
\ell(\bm{\eta}) = \sum\limits_{t=2}^{T}\log\left[\textrm{Pr}(X_{t}=k|X_{t-1}=l)\right],
\end{equation*}
with $\textrm{Pr}(X_{t}=k|X_{t-1}=l)$ as in (\ref{cmldp}) or (\ref{cmlgp}).
CML estimates $ \widehat{\bm{\eta}}_{\textrm{CML}}$ for $\bm{\eta}$ are obtained by maximizing $\ell(\bm{\eta})$.
In practical scenery there will be no closed form for the CML estimates and numerical methods need to be necessary. As starting values for the algorithm, we have used the estimates obtained by the YW or CLS methods. Since the Fisher information matrix is not available, the standard errors are obtained as the square roots of the elements in
the diagonal of the inverse of the negative of the Hessian of the conditional log-likelihood
calculated at the CML estimates.
\end{remark}

Tables \ref{tabsim1} and \ref{tabsim2} show the empirical bias and mean square error of the estimators obtained from the YW, CLS and CML methods under \textbf{INARDP}(1) and \textbf{INARGP}(1) models, respectively. The results set out in Table~\ref{tabsim1}, under \textbf{INARDP}(1) model, show that the CML estimators has the best performance on empirical bias and MSE compared with the YW and CLS estimators. For the estimators of $\alpha$ and different values of $\phi$, we notice that the MSE of both methods are very much similar.  Note that for large sample size both methods given a smaller bias and MSE (very close to zero) for estimates of $\alpha, \mu $ and $\phi$.  The bias of the estimators of $\alpha$ are negative and does not depend on the values of other parameters considered.

However, the results displayed in Table ~\ref{tabsim2}, under \textbf{INARGP}(1) model, reveal that for the overdispersed case the estimator $\widehat{\alpha}_{\textrm{CML}}$ turns out to be better than the other methods as it gives in all cases lower bias (MSE) than the YW and CLS methods. However,
for the underdispersed case we have a reverse scenery. For example, for the estimador of $\alpha$, the CLS method has the best performance on empirical bias. However, when estimating $\phi$, the CML method shows the best performance on empirical bias. Finally, we notice that the CLS and YW methods present similar MSE behaviors.

\begin{table}[t]
\small
\caption{Empirical bias and MSE (in parentheses) of estimators of $\alpha, \mu, \phi$.}\label{tabsim1}
\renewcommand{\arraystretch}{1.2}
\resizebox{\linewidth}{!}{
\begin{tabular}{lrrrrrrrrrrr}
 \hline
\multirow{2}{*}{$T$}&\multicolumn{3}{c}{Estimator of $\alpha$}&&\multicolumn{3}{c}{Estimator of $\mu$}&&\multicolumn{3}{c}{Estimator of $\phi$}\\ \cline{2-4} \cline{6-8} \cline{10-12}
& \multicolumn{1}{c}{$\widehat{\alpha}_{\textrm{CLS}}$}&\multicolumn{1}{c}{$\widehat{\alpha}_{\textrm{YW}}$}&\multicolumn{1}{c}{$\widehat{\alpha}_{\textrm{CML}}$}&& \multicolumn{1}{c}{$\widehat{\mu}_{\textrm{CLS}}$}&\multicolumn{1}{c}{$\widehat{\mu}_{\textrm{YW}}$}&\multicolumn{1}{c}{$\widehat{\mu}_{\textrm{CML}}$}&& $\widehat{\phi}_{\textrm{CLS}}$&$\widehat{\phi}_{\textrm{YW}}$&$\widehat{\phi}_{\textrm{CML}}$\\ \hline
& \multicolumn{11}{c}{$\alpha = 0.3, \mu = 5.0$ and $\phi = 0.5$ (overdispersed case)}\\ \cline{2-12}
\multirow{2}{*}{100} &$-$0.0188 & $-$0.0218 &$-$0.0168& &0.0855 &0.1060&0.0998& &0.0302 & 0.0246&0.0245 \\
    &(0.0101) &(0.0101) &(0.0090) & &(0.5900) &(0.5842)&(0.5502)& &(0.0119) &(0.0112) &(0.0123) \\
\multirow{2}{*}{200} & $-$0.0103 &$-$0.0118 &$-$0.0079& &0.0295 &0.0398&0.0493& &0.0141 &0.0114 &0.0107 \\
    &(0.0048) &(0.0048) &(0.0041) & &(0.2766) &(0.2754)&(0.2489)& &(0.0052) &(0.0050) &(0.0053) \\
\multirow{2}{*}{400} & $-$0.0051&$-$0.0059 &$-$0.0043 & &$-$0.0105 &$-$0.0053&0.0236& &0.0069 &0.0055 &0.0056 \\
    &(0.0023) &(0.0023) &(0.0019) & &(0.1365) &(0.1356)&(0.1159)& &(0.0025) &(0.0025) &(0.0026) \\
\multirow{2}{*}{800} &$-$0.0017 &$-$0.0020 &$-$0.0013& &$-$0.0324 &$-$0.0298&0.0063& &0.0022 &0.0015 &0.0020 \\
    &(0.0012) &(0.0012) &(0.0010) & &(0.0704) &(0.0701)&(0.0585)& &(0.0012) &(0.0012) &(0.0012) \\
\hline
& \multicolumn{11}{c}{$\alpha = 0.3, \mu = 5.0$ and $\phi = 2.0$ (underdispersed case)}\\ \cline{2-12}
\multirow{2}{*}{100} &$-$0.0193 &$-$0.0223 &$-$0.0185& &0.1399 &0.1613&0.1289& &0.1348 &0.0926 &0.1263 \\
    &(0.0097) &(0.0096) &(0.0094) & &(0.5002) &(0.4974)&(0.4865)& &(0.2972) &(0.2649) &(0.2800) \\
\multirow{2}{*}{200} &$-$0.0110 &$-$0.0124 &$-$0.0105& & 0.0814&0.0916&0.0729& &0.0671 &0.0474 &0.0622 \\
    &(0.0048) &(0.0048) &(0.0046) & &(0.2521) &(0.2518)&(0.2417)& &(0.1199) &(0.1134) &(0.1135) \\
\multirow{2}{*}{400} &$-$0.0054 &$-$0.0061 &$-$0.0051& &0.0421 &0.0472&0.0344& &0.0299 &0.0204 &0.0240 \\
    &(0.0024) &(0.0024) &(0.0023) & &(0.1265) &(0.1264)&(0.1217)& &(0.0549) &(0.0535) &(0.0508) \\
\multirow{2}{*}{800} & $-$0.0021& $-$0.0025 &$-$0.0021& &0.0191 &0.0217&0.0134& &0.0175 &0.0128 &0.0125 \\
    &(0.0012) &(0.0012) &(0.0012) & &(0.0631) &(0.0631)&(0.0606)& &(0.0276) &(0.0272) &(0.0259) \\
\hline
\end{tabular}
}
\bigskip
\end{table}

\begin{table}[t]
\centering
\small
\caption{Empirical bias and MSE (in parentheses) of estimators of $\alpha, \mu, \phi$.}\label{tabsim2}
\renewcommand{\arraystretch}{1.2}
\resizebox{\linewidth}{!}{
\begin{tabular}{lrrrrrrrrrr}
\hline
\multirow{2}{*}{$T$}&\multicolumn{3}{c}{Estimator of $\alpha$}&&\multicolumn{2}{c}{Estimator of $\mu$}&&\multicolumn{3}{c}{Estimator of $\phi$}\\ \cline{2-4} \cline{6-7} \cline{9-11}
& \multicolumn{1}{c}{$\widehat{\alpha}_{\textrm{CLS}}$}&\multicolumn{1}{c}{$\widehat{\alpha}_{\textrm{YW}}$}&\multicolumn{1}{c}{$\widehat{\alpha}_{\textrm{CML}}$}&& \multicolumn{1}{c}{$\widehat{\mu}_{\textrm{YW}}$}&\multicolumn{1}{c}{$\widehat{\mu}_{\textrm{CML}}$}&& \multicolumn{1}{c}{$\widehat{\phi}_{\textrm{CLS}}$}&\multicolumn{1}{c}{$\widehat{\phi}_{\textrm{YW}}$}&\multicolumn{1}{c}{$\widehat{\phi}_{\textrm{CML}}$}\\ \hline
& \multicolumn{10}{c}{$\alpha = 0.3, \mu = 1.0$ and $\phi = -0.5$  (underdispersed case)}\\ \cline{2-11}
\multirow{2}{*}{100} & $-$0.0263 &$-$0.0289 &0.1994& &$-$0.0461 &$-$0.3057& & 0.1458&0.1524 &$-$0.0785 \\
    &(0.0116) &(0.0116) &(0.0977) & &(0.0413) &(0.1580)& &(0.0503) &(0.0518) &(0.1509) \\
\multirow{2}{*}{200} &$-$0.0182 & $-$0.0197&0.1943& &$-$0.0659 &$-$0.3141& &0.1520 &0.1552 &$-$0.0465 \\
    &(0.0063) &(0.0063) &(0.0860) &  &(0.0255)&(0.1364)& &(0.0433) &(0.0441) &(0.1397) \\
\multirow{2}{*}{400} & $-$0.0047&$-$0.0054  &0.1034& &$-$0.1054 &$-$0.2483 & &0.2057 &0.2073 & 0.1300\\
    &(0.0032 ) &(0.0032) &(0.0429) &  &(0.0211)&(0.0840)& &(0.0508) &(0.0514) &(0.1172) \\
\multirow{2}{*}{800} & $-$0.0037 &$-$0.0041 &0.0748& &$-$0.1102 &$-$0.2296&&0.2134 &0.2142  & 0.1902\\
    &(0.0017) &(0.0017) &(0.0285) &  &(0.0175)&(0.0675)& &(0.0511) &(0.0514) &(0.1038) \\
\hline
& \multicolumn{10}{c}{$\alpha = 0.3, \mu = 1.0$ and $\phi = 0.5$ (overdispersed case)}\\ \cline{2-11}
\multirow{2}{*}{100} &$-$0.0184 & $-$0.0214  &$-$0.0071& &0.0788 &0.0293&& $-$0.0284 & $-$0.0259&$-$0.0146 \\
    &(0.0100) &(0.0099) &(0.0040) &  &(0.0704)&(0.0368)& &(0.0069) &(0.0067) &(0.0055) \\
\multirow{2}{*}{200} & $-$0.0108&$-$0.0123 &$-$0.0042& &0.0437 &0.0158&&$-$0.0143  & $-$0.0133 & $-$0.0066 \\
    &(0.0049) &(0.0049) &(0.0019) &  &(0.0335)&(0.0165)& &(0.0033) &(0.0033) &(0.0024) \\
\multirow{2}{*}{400} & $-$0.0044&$-$0.0051 &$-$0.0016& &0.0221 &0.0075&&$-$0.0083 & $-$0.0078 &$-$0.0037 \\
    &(0.0025) &(0.0025) &(0.0009) &  &(0.0160)&(0.0075)& &(0.0017) &(0.0017) &(0.0012) \\
\multirow{2}{*}{800} &$-$0.0030 & $-$0.0034&$-$0.0012& & 0.0122&0.0050&&$-$0.0040 & $-$0.0037 &$-$0.0020 \\
    &(0.0013) &(0.0013) &(0.0005) &  &(0.0081)&(0.0039)& &(0.0009) &(0.0009) &(0.0006) \\
\hline
\end{tabular}
}
\bigskip
\end{table}

\section{Real data examples}
\hskip 0.65cm

To illustrate the applications of the proposed models, we consider in this section two real data sets with
overdispersion and underdispersion. We compared the proposed
processes with the \textrm{INARP}(1) process (special case). In order
to estimate the parameters of these processes, we adopt the CML method and all the computations were done using the {\tt tsinteger} package.

\begin{remark}(Detecting overdispersion or underdispersion)
For testing the null hypothesis  $\mathcal{H}_0: X_1, \ldots,  X_T$ stem from an equidispersed Poisson \textrm{INAR}(1)
process $(\textrm{FI}_X = 1)$ against the alternative of an overdispersed (or underdispersed) marginal distribution,
we suggest to use the the following test provided by \cite{Schweer:2014aa}. Let $z_{1-\beta}$ be the quantile of the $(1 - \beta)$-quantile of the N(0,1)-distribution, that is, $\Phi(z_{1-\beta}) = 1 - \beta$, for $\beta \in (0, 1)$, where $\Phi(\cdot)$ is the distribution function of
a N(0,1)-distribution. We reject the null hypothesis $\mathcal{H}_0: \phi = 1$ or $\phi = 0$ (equidispersion) in favor of alternative hypothesis
$\mathcal{H}_1: \phi < 1$ or $\phi > 0$ (overdispersion) if
\begin{equation*}\label{test1}
\widehat{\textrm{FI}}_X > z_{1-\beta}\sqrt{\frac{2(1 + \alpha^2)}{T(1-\alpha^2)}},
\end{equation*}
where $\widehat{\textrm{FI}}_X := \sum_{t=1}^{T}(X_t- \overline{X})^2/\sum_{t=1}^{T}X_t$ with $\overline{X} := (1/T)\sum_{t=1}^{T}X_t$.
Furthermore, if the alternative hypothesis of interest is $\mathcal{H}_1: \phi > 1$ or $\phi < 0$ (underdispersion), we reject $\mathcal{H}_0$ in favour of an alternative hypothesis if
\begin{equation*}\label{test2}
\widehat{\textrm{FI}}_X < z_{\beta}\sqrt{\frac{2(1 + \alpha^2)}{T(1-\alpha^2)}}.
\end{equation*}
\end{remark}

\subsection{Overdispersed data: weekly number of syphilis cases}
\hskip 0.65cm

As a first example, we consider the data set consisting of the weekly number of syphilis cases in the United States from 2007 to 2010 in Mid-Atlantic states given in {\tt tsinteger} package available for download at {\tt data(syphillis)}. The data consist of 209 observations, and they were already analyzed by \cite{Borges:2017aa}.

The sample mean is 24.63, the sample variance is 105.68, and the first-order autocorrelation is 0.2322. The empirical Fisher index of dispersion is 4.29. The sample variance is much larger than the sample mean, hence, the data seems to be overdispersed. The equidispersion test \citep[][]{Schweer:2014aa} rejected the null hypothesis of equidispersion, the $p$-value for the test being $ < 0.01$. Consequently, a Poisson marginal distribution seems to not be appropriate.

\begin{figure}[!h]
\centering
\includegraphics[width=0.8\textwidth]{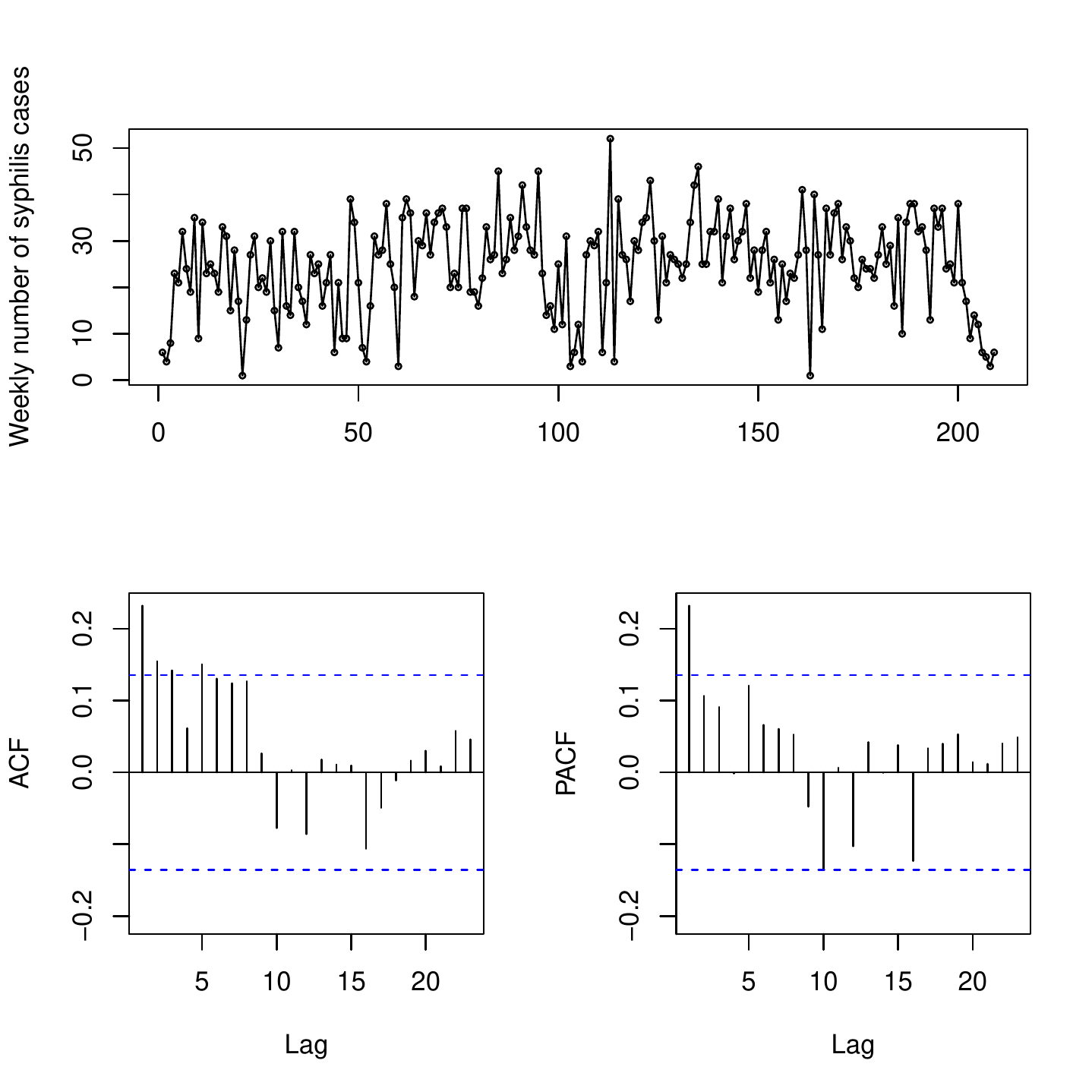}
\caption{Plots of the time series, autocorrelation and partial autocorrelation functions for the number of syphilis cases.}
\label{fig_aplic1}
\end{figure}

The series together with its sample autocorrelation and partial autocorrelation functions are displayed in Figure \ref{fig_aplic1}. Analysing the Figure \ref{fig_aplic1} we conclude that a first-order autoregressive model may be appropriate for the given data series, given the pattern of the sample partial autocorrelation function and the clear cut-off.

Table \ref{tableaplic1} gives the CML estimates (with corresponding standard errors in parentheses), Akaike information criterion (AIC) and Bayesian
information criterion (BIC) for the fitted models. Since the values of the AIC and BIC are smaller for the \textbf{INARGP}(1) and \textbf{INARDP}(1) models compared to those values of the \textbf{INARP}(1) model. The likelihood ratio (LR) statistic to test the hypothesis $\mathcal{H}_0:$ \textbf{INARP}(1) against the alternative hypothesis $\mathcal{H}_1:$ \textbf{INARGP}(1), i.e., $\mathcal{H}_0: \phi = 0$ against $\mathcal{H}_1: \phi \neq 0$, is 403.39 ($p$-value $< 0.01$). Furthermore, the LR statistic to test the hypothesis $\mathcal{H}_0:$ \textbf{INARP}(1) against the alternative hypothesis $\mathcal{H}_1:$ \textbf{INARDP}(1), i.e., $\mathcal{H}_0: \phi = 1$ against $\mathcal{H}_1: \phi \neq 1$, is 453.04 ($p$-value $< 0.01$).
Thus, we reject the null hypothesis in favor of the \textbf{INARGP}(1) and \textbf{INARDP}(1) models using any usual significance level.
Therefore, the \textbf{INARGP}(1) and \textbf{INARDP}(1) models are significantly better than the \textbf{INARP}(1) model based on the LR statistic.

\begin{table}[!htbp]
\centering
\caption{Estimates of the parameters , MSE (in parentheses), AIC, BIC and estimated quantities
for the number of syphilis cases.}\label{tableaplic1}
\renewcommand{\arraystretch}{1.3}
\resizebox{\linewidth}{!}{
\begin{tabular}{lcccccccc}
 \hline
Model            &Parameter &CML Estimate            &AIC      &BIC  &&$\mu_X$& $\sigma^2_X$& $\textrm{FI}_X$    \\
 \hline
\multirow{3}{*}{\textrm{INARGP}(1)}        &$\alpha$  &  $0.0798 \, (0.0497)$       &\multirow{3}{*}{1615.15}         &\multirow{3}{*}{1625.18}     &&\multirow{3}{*}{24.72}       &\multirow{3}{*}{137.04}            &\multirow{3}{*}{5.54}        \\
                 &$\mu$ &  $9.3614 \, (0.8164)$       &         &     &&       &            &   \\
                 &$\phi$  &  $0.5885 \, (0.0255)$       &         &     &&       &            &   \\\\
\multirow{3}{*}{\textrm{INARDP}(1)}        &$\alpha$  &  $0.1154 \, (0.0404) $       &\multirow{3}{*}{1565.50}         &\multirow{3}{*}{1575.53}     &&\multirow{3}{*}{24.84}       &\multirow{3}{*}{113.89}            &\multirow{3}{*}{4.58}  \\
                 &$\mu$     &  $21.976 \, (1.2204) $       &         &     &&       &            &   \\
                 &$\phi$    & $ 0.2001 \, (0.0195)$        &         &     &&       &            &   \\\\
\multirow{2}{*}{\textrm{INARP}(1)}         &$\alpha$  &$ 0.1480 \, (0.0261) $       &\multirow{2}{*}{2016.54}         &\multirow{2}{*}{2023.22}     &&\multirow{2}{*}{24.72}       &\multirow{2}{*}{24.72}            & \multirow{2}{*}{1} \\
                 &$\mu$ &$  21.063 \, (0.7087)$       &         &     &&       &            &   \\
\hline
Empirical        &         &                &         &     &&24.63  &105.68    &4.29  \\ \hline
\end{tabular}
}
\bigskip
\end{table}

We note that $\widehat{\phi}_{\mathrm{CML}} > 0$ and $\widehat{\phi}_{\mathrm{CML}} < 1$ for the \textbf{INARGP}(1) and \textbf{INARDP}(1) models, respectively,
which implies that the dispersion index of these models is greater than 1
in accordance with equidispersion test~\citep[][]{Schweer:2014aa}. From the figures of Table \ref{tableaplic1} and
according to LR tests, the \textbf{INARDP}(1) model fits the current
data better than other models, i.e., these values indicate that the
null hypothesis is strongly rejected for the \textbf{INARP}(1) model.
These results illustrate the potentiality of the \textbf{INARGP}(1) and \textbf{INARDP}(1) models
and the importance of the additional parameter [in \textbf{INARP}(1) model]. Also, the residuals are not
correlated.

\subsection{Underdispersed data: family violence counts}
\hskip 0.65cm

As the second example, we consider the series of monthly counts of family violences
in the 11th police car beat in Pittsburgh, during one month.
It consists of 143 observations, starting in January 1990 and ending in November 2001.
The data set is obtained from {\tt tsinteger} package by {\tt data(violences)}.

\begin{figure}[!h]
\centering
		 \includegraphics[width=0.8\textwidth]{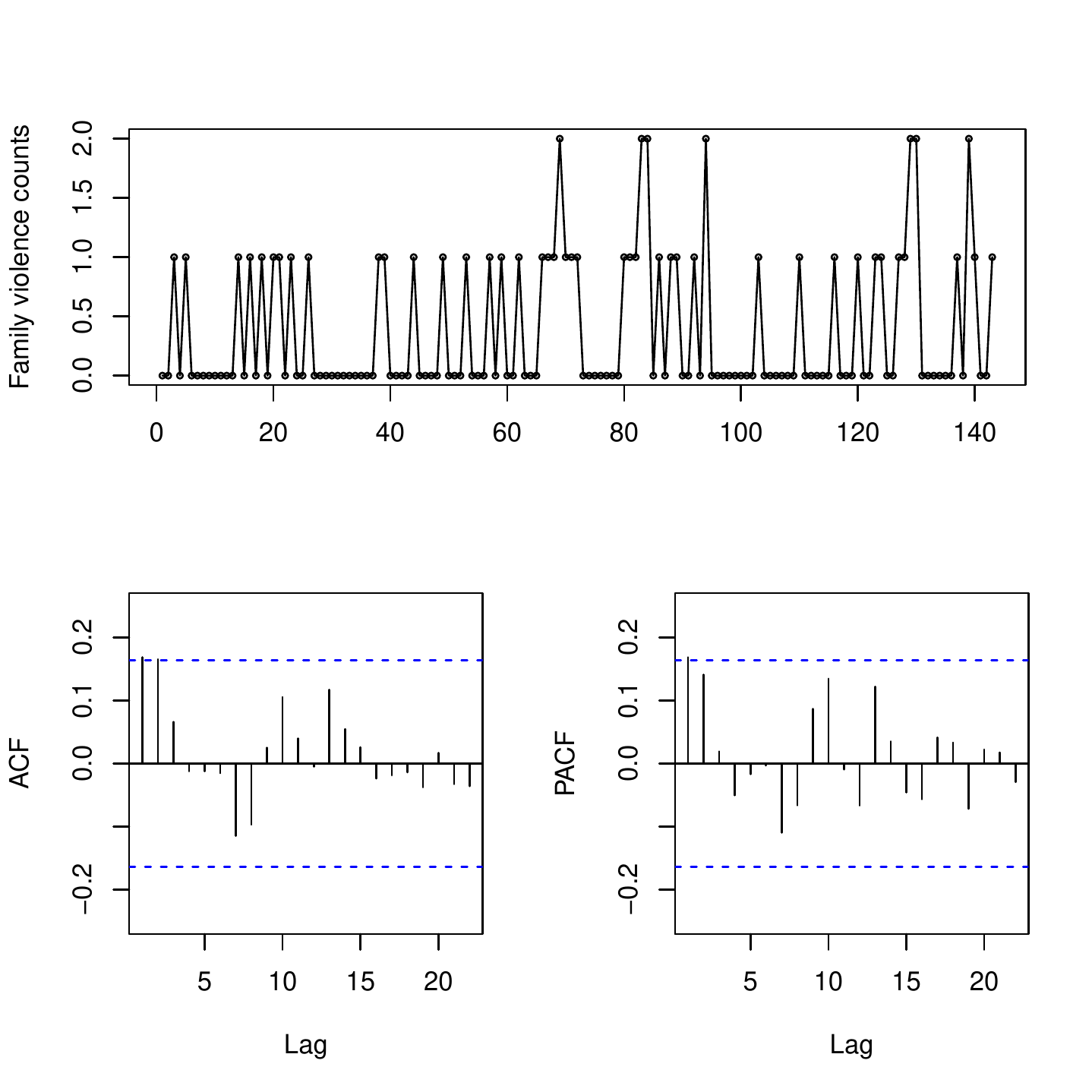}
\centering
\caption{Plots of the time series, autocorrelation and partial autocorrelation functions for the family violence counts.}
\label{fig_aplic2}
\end{figure}

The sample mean, variance and Fisher index of dispersion are 0.3846, 0.3369 and 0.8761, respectively,
which indicates that the data are underdispersed. The first-order autocorrelation is 0.166.
The series together with its sample autocorrelation and partial autocorrelation functions
is displayed in Figure \ref{fig_aplic2}.

Analyzing Figure \ref{fig_aplic2}, we conclude that the first order autoregressive models may
be appropriate for the given data series. The behavior of the series indicates that it
may be a mean stationary time series. So, we apply the \textbf{INARGP}(1), \textbf{INARDP}(1) and \textbf{INARP}(1) models to the data. Parameter
estimates and their standard errors are summarized in Table \ref{tableaplic2}. The AIC and BIC values are also provided.

\begin{table}[!htbp]
\centering
\caption{Estimates of the parameters (MSE in parentheses), AIC, BIC and estimated quantities
for the family violence counts.}\label{tableaplic2}
\renewcommand{\arraystretch}{1.3}
\resizebox{\linewidth}{!}{
\begin{tabular}{lcrcccccc}
 \hline
Model            &Parameter &CML Estimate            &AIC      &BIC  &&$\mu_X$& $\sigma^2_X$& $\textrm{FI}_X$    \\
 \hline
\multirow{3}{*}{\textrm{INARGP}(1)}        &$\alpha$  &$0.1613 \, (0.0833)$       & \multirow{3}{*}{223.86}        &\multirow{3}{*}{232.75}     && \multirow{3}{*}{0.3887}      &\multirow{3}{*}{0.3236}            &\multirow{3}{*}{0.8325}        \\
                 &$\mu$ &$0.3632 \, (0.0627)$        &         &     &&       &            &   \\
                 &$\phi$  &$-0.1142 \, (0.0527)$       &         &     &&       &            &   \\\\
\multirow{3}{*}{\textrm{INARDP}(1)}        &$\alpha$  &  $0.1924 \, (0.0893) $       &\multirow{3}{*}{223.64}         &\multirow{3}{*}{232.53}     &&\multirow{3}{*}{0.3890}       &\multirow{3}{*}{0.3204}            &\multirow{3}{*}{0.8236}  \\
                 &$\mu$     &  $0.3141 \, (0.0498) $       &         &     &&       &            &   \\
                 &$\phi$    &  $1.2664 \, (0.1576) $       &         &     &&       &            &   \\\\
\multirow{2}{*}{\textrm{INARP}(1)}         &$\alpha$  &$ 0.1562 \, (0.0931) $       &\multirow{2}{*}{224.98}         &\multirow{2}{*}{230.91}     &&\multirow{2}{*}{0.3886}       &\multirow{2}{*}{0.3886}            & \multirow{2}{*}{1} \\
                 &$\mu$ &$ 0.3279 \, (0.0566) $       &         &     &&       &            &   \\
\hline
Empirical        &         &                &         &     &&0.3846  & 0.3369    &0.8761  \\ \hline
\end{tabular}
}
\bigskip
\end{table}

Analyzing Table \ref{tableaplic2}, note that $\widehat{\phi}_{\mathrm{CML}} < 0$ and $\widehat{\phi}_{\mathrm{CML}} > 1$ for the \textbf{INARGP}(1) and \textbf{INARDP}(1) models, respectively, which implies that the dispersion index of these models is less than 1.
Based on AIC, we find that the \textbf{INARGP}(1) and \textbf{INARDP}(1) models are the best ones. Based on BIC, we find
that the \textbf{INARP}(1) is the best one. Within these three fitted models, the mean,
variance and dispersion index are summarized in Table \ref{tableaplic2}. All three models exhibit good fits of mean, but only the \textbf{INARGP}(1) and \textbf{INARDP}(1) models give a reasonable fit of variance and present underdispersed features. As a result, the \textbf{INARP}(1) model is not adequate for the data, e. g., its variance 0.3886 is much large than the empirical variance 0.3369, and also its dispersion index (1 vs. 0.8761). Based on this fact, we
conclude that the \textbf{INARGP}(1) and \textbf{INARDP}(1) models capture more information of these data.

We test the null hypothesis $\mathcal{H}_0:$ \textbf{INARP}(1) against the alternative hypothesis $\mathcal{H}_1:$ \textbf{INARGP}(1), i.e., $\mathcal{H}_0: \phi = 0$ against $\mathcal{H}_1: \phi \neq 0$ (with a significance level at $10\%$). The LR statistic to test the hypothesis is 3.123 ($p$-value is 0.0772).
Furthermore, the LR statistic to test the hypothesis $\mathcal{H}_0:$ \textbf{INARP}(1) against the alternative hypothesis $\mathcal{H}_1:$ \textbf{INARDP}(1), i.e., $\mathcal{H}_0: \phi = 1$ against $\mathcal{H}_1: \phi \neq 1$, is 3.342 ($p$-value is 0.0675).
Thus, we reject the null hypothesis in favor of the \textbf{INARGP}(1) and \textbf{INARDP}(1) models.
Therefore, the \textbf{INARGP}(1) and \textbf{INARDP}(1) models are significantly better than the \textbf{INARP}(1) model based on the LR statistic.

\section*{Appendix}
\hskip 0.65cm

\begin{appendix}
\begin{proof}{Proposition 1:}\label{proof1}

The proof of Proposition 1 is omitted here since it is a straightforward consequence of an application
obtained by \cite{Klimko:1978aa},  p. 638.
\end{proof}

\begin{proof}{Proposition 2:}
%\added[id=Manoel]{Proof }{}

Let $X_1, \ldots, X_T$ be a sample of an INARGP(1) process.
It can be verified that the regularity conditions given in Theorem 3.2 of \cite{Klimko:1978aa}, p. 634, are satisfied by
INARGP(1) process.

%From \cite{Barreto-Souza:2015ab} we have following fact: the estimators $\widehat{\alpha}_{\textrm{CLS}}$ and $\widehat{\mu}_{\textrm{CLS}}$ given in (\ref{cls1}) are %strongly consistent and satisfy the asymptotic normality
%$$\sqrt{T}[(\widehat{\alpha}_{\textrm{CLS}},\widehat{\phi}_{\textrm{CLS}})^\top-(\alpha,\phi)^\top]\stackrel{d}{\longrightarrow} \textrm{N}_2(\bm{0}^\top,
%\bm{i(\eta)}^{-1}),$$
%where $\bm{i(\eta)}^{-1} = \bm{V}^{-1}\bm{W}\bm{V}^{-\top}$.

Consider the following quantities $\textrm{E}_{t|t-1} \equiv \textrm{E}(X_t|X_{t-1}) = \alpha X_{t-1} + \mu/(1-\phi)$ and $\textrm{d}_{t|t-1} = \textrm{Var}(X_t|X_{t-1}) =  \alpha(1-\alpha)X_{t-1} + \mu/(1-\phi)^3$, and calculate
\begin{eqnarray*}
\frac{\partial \textrm{E}_{t|t-1}}{\partial \alpha}=X_{t-1}, \quad
\frac{\partial \textrm{E}_{t|t-1}}{\partial \phi}  = \frac{\mu}{(1-\phi)^2}, \quad
\frac{\partial^2 \textrm{E}_{t|t-1}}{\partial \alpha^2}  = 0, \quad
\frac{\partial^2 \textrm{E}_{t|t-1}}{\partial \phi^2}  = \frac{2\mu}{(1-\phi)^3},\quad
\frac{\partial^2 \textrm{E}_{t|t-1}}{\partial \phi \partial \alpha}  = 0.
\end{eqnarray*}
%

%Finally, the matrices $\bm{V}$ and $\bm{W}$ were obtained as follows
Define the $2 \times 2$ matrix $\bm{V}$ according to Equation (3.2) in \cite{Klimko:1978aa} as

\[
\bm{V} = \textrm{E}\left(
\begin{bmatrix}
\frac{\partial \textrm{E}_{t|t-1}}{\partial \alpha} \\
\frac{\partial \textrm{E}_{t|t-1}}{\partial \phi}
\end{bmatrix}
\begin{bmatrix}
\frac{\partial \textrm{E}_{t|t-1}}{\partial \alpha} &
\frac{\partial \textrm{E}_{t|t-1}}{\partial \phi}
\end{bmatrix}
\right)
=
\begin{pmatrix}
\textrm{E}(X_{t-1}^2)&  \frac{\mu}{(1-\phi)^2}\textrm{E}(X_{t-1})  \\
 \frac{\mu}{(1-\phi)^2}\textrm{E}(X_{t-1})  & \frac{\mu^2}{(1-\phi)^4}
\end{pmatrix} =
\begin{pmatrix}
\mu_{X,2} & (1-\alpha)\mu_x\\
(1-\alpha)\mu_x &  \mu_\epsilon \sigma^2_\epsilon
\end{pmatrix}
\]
and the $2 \times 2$ matrix $\bm{W}$ according to Equation (3.5) in \cite{Klimko:1978aa} as

\[
\bm{W} =  \textrm{E}\left(
\begin{bmatrix}
\frac{\partial \textrm{E}_{t|t-1}}{\partial \alpha} \\
\frac{\partial \textrm{E}_{t|t-1}}{\partial \phi}
\end{bmatrix}
\textrm{d}_{t|t-1}
\begin{bmatrix}
\frac{\partial \textrm{E}_{t|t-1}}{\partial \alpha} &
\frac{\partial \textrm{E}_{t|t-1}}{\partial \phi}
\end{bmatrix}
\right)
=
\left(
\begin{array}{ccccc}
\alpha(1-\alpha)\mu_{X,3} + \sigma_{\epsilon}^{2}\mu_{X,2} &&&& \mu_{\epsilon}\,\sigma_{\epsilon}^{2}(\alpha\,\mu_{\epsilon} + \sigma_{\epsilon}^{2}) \\ \\ \\
\mu_{\epsilon}\,\sigma_{\epsilon}^{2}(\alpha\,\mu_{\epsilon} + \sigma_{\epsilon}^{2})&&&&
\lambda\left[\alpha(1-\alpha)\mu_{X,2} + \mu_X\sigma_{\epsilon}^{2}\right]/(1-\theta)^2
\end{array}
\right).
\]

Hence, the estimators $\widehat{\alpha}_{\textrm{CLS}}$ and $\widehat{\phi}_{\textrm{CLS}}$ CLS of $\alpha$ and $\phi$ have the following asymptotic distribution:

$$\sqrt{T}[(\widehat{\alpha}_{\textrm{CLS}},\widehat{\phi}_{\textrm{CLS}})^\top-(\alpha,\phi)^\top]\stackrel{d}{\longrightarrow}
\textrm{N}_2((0, 0)^\top,
\bm{V}^{-1}\bm{W}\bm{V}^{-1}).$$

%The inverse of
%$\bm{V}$ is given by
%\[
%\bm{V}^{-1} = \begin{pmatrix}
%\frac{\mu_{\epsilon}\sigma_{\epsilon}^{2}}{\mu_{\epsilon}\sigma_{\epsilon}^{2}\mu_{X,2} - (1-\alpha)^2\mu_{X}^4} &-\frac{(1-\alpha)\mu_{X}^2}{\mu_{\epsilon}\sigma_{\epsilon}^{2}\mu_{X,2} - (1-\alpha)^2\mu_{X}^4}  \\
%& \\
%-\frac{(1-\alpha)\mu_{X}^2}{\mu_{\epsilon}\sigma_{\epsilon}^{2}\mu_{X,2} - (1-\alpha)^2\mu_{X}^4}&
%\frac{\mu_{X,2}}{\mu_{\epsilon}\sigma_{\epsilon}^{2}\mu_{X,2} - (1-\alpha)^2\mu_{X}^4}
%\end{pmatrix}.
%\]
\end{proof}

\end{appendix}

\end{document}